\documentstyle[11pt,aaspp4]{article}

\def\etal{{\it et al.}}
\def\solar{$_{\sun}$}

\lefthead{Looney et al.}
\righthead{L1551~IRS5: A protobinary system?}
\begin{document}

\title{High Resolution $\lambda$ = 2.7 mm Observations of L1551~IRS5:\\ 
A Protobinary System?}
\author{Leslie W. Looney\altaffilmark{1}, Lee G. Mundy\altaffilmark{2}}
\affil{Department of Astronomy, University of Maryland, College Park}

\and

\author{W. J. Welch\altaffilmark{3}}
\affil{Radio Astronomy Laboratory, University of California, Berkeley}

\altaffiltext{1}{Email: lwl@astro.umd.edu}
\altaffiltext{2}{Email: lgm@astro.umd.edu}
\altaffiltext{3}{Email: welch@jack.berkeley.edu}

\begin{abstract}

We present sub-arcsecond resolution imaging of the $\lambda$~=~2.7~mm
continuum emission from the young, embedded
system L1551~IRS5 using the nine-element,
high-resolution configuration of the BIMA array.
The observed emission arises from
two compact sources separated by $0\farcs$35,
coinciding with the two sources seen at
$\lambda$~=~2~cm and $\lambda$~=~1.3~cm.
When the high resolution data is combined with data 
from two compact configurations,
L1551~IRS5 is argued to consist of a 
protobinary system separated by 
$\sim$50~AU with individual circumstellar disks, a
circumbinary structure, and a large-scale envelope.  
The characteristic masses of the components are:
0.024~M$_{\sun}$ for the northern circumstellar disk,
0.009~M$_{\sun}$ for the southern circumstellar disk,
0.04~M$_{\sun}$ for the circumbinary material, 
and 0.28~M$_{\sun}$ for the envelope.

\end{abstract}


\section{Introduction}

First detected in an infrared survey of the L1551 cloud\markcite{ssv}
(Strom, Strom, \& Vrba 1976), L1551~IRS5 is a prototypical 
young stellar system, with a strong bipolar molecular
outflow\markcite{snell}
(Snell, Loren, \& Plambeck 1980), an optical jet\markcite{mundt}
(Mundt \& Fried 1983), HH objects\markcite{herbig}
(Herbig 1974), and an envelope-disk structure in the
surrounding material\markcite{kandm} (Keene \& Masson 1990).
Located at a distance of 140 pc\markcite{elias} (Elias 1978) and exhibiting
a luminosity of $\sim 28 L_{\sun}$\markcite{butner} (Butner \etal\  1991),
L1551~IRS5 was one of the defining examples for  
Class I sources in the classification
scheme of\markcite{adams} Adams, Lada, and Shu (1987)
and has been used as an archetype in the current
paradigm for single-star formation\markcite{shu93} (Shu \etal\ 1993).
But is it really a single-star system?

High resolution $\lambda$~=~2~cm continuum observations of L1551~IRS5
show two compact sources with a separation of
$\sim 0\farcs 28$\markcite{bieging}\markcite{rod86}
(Bieging \& Cohen 1985; Rodr\'iguez \etal\ 1986)
which have been interpreted as either a protobinary system 
\markcite{bieging}(Bieging \& Cohen 1985),
or the inner ionized edges of a gas and dust toroid surrounding a
single star\markcite{rod86} (Rodr\'iguez \etal\ 1986). 
The latter is the
most widely accepted interpretation, but
comparisons with $\lambda$~=~2~cm emission from
other young
binary systems such as T Tau and Z CMa\markcite{bcs} 
\markcite{schwartz}(Bieging, Cohen, \& Schwartz 1984; 
Schwartz, Simon, \& Zuckerman 1983),
suggest that the binary interpretation is also viable.

Under the assumption that L1551~IRS5 is a single star system, 
Keene and Masson\markcite{kandm90} (1990) modeled $\lambda$~=~2.7~mm
interferometric observations to deduce the presence of
a 45~AU radius circumstellar disk within an envelope.
This envelope, which extends out $\sim$1000~AU from IRS5,
contains 0.1 to 1 M\solar\  of material
\markcite{ladd95}\markcite{fuller95}
(Ladd \etal\ 1995; Fuller \etal\ 1995).
High resolution JCMT-CSO interferometric observations at $\lambda$~=~870~$\mu$m 
resolved the compact central emission (\markcite{lay}Lay \etal\ 1994),
and the emission was modeled as arising from an 80~AU radius
Gaussian source,
inferred to be an accretion disk around the young star.

In this letter, we present sub-arcsecond imaging 
of the $\lambda$~=~2.7~mm continuum emission from 
the L1551~IRS5 system.
These observations re-open questions about the binarity of 
the system and the distribution of the surrounding material.

\section{Observations and Data Reduction}

L1551~IRS5 was observed in three array configurations of the 
9-element BIMA Array\footnotemark \markcite{welch} (Welch \etal\ 1996).
The longest baselines were 1~km N-S and 900 m E-W, yielding a
maximum projected baseline of 480~k$\lambda$ (1.4km); the shortest 
baselines were limited by the antenna size of 6.1~m, yielding
a minimum projected baseline of 2.2~k$\lambda$.
This range in projected baselines provides images with a minimum
resolution of 0$\farcs$3, fully sampled to
sizes as large as 60$\arcsec$.

\footnotetext{ The BIMA Array is operated by the Berkeley Illinois Maryland
Association under funding from the National Science Foundation.}

For the high resolution configuration (March 1, 1996), atmospheric phase 
fluctuations were tracked by switching the antennas
between source, phase calibrator, and a nearby weak quasar on a 
two minute cycle.
The usefulness of this quick switching technique has been 
demonstrated at the VLA (\markcite{holdaway}Holdaway \& Owen 1995).
The main phase calibrator (0530+135) was used to track rapid
atmospheric phase fluctuations.  The secondary quasar 
(0449+113) was used to track slow phase drifts due to the 
difference in airmass
between the primary calibrator and source and, more importantly
for this array, phase drifts due to uncertainties in baseline length.

The digital correlator was configured with two 700~MHz bands 
centered at 107~GHz and 109~GHz. 
The flux amplitude calibration assumed a flux of 6.8~Jy for 
0530+135, as observed in the following month's compact array.
The coherence of the atmosphere was checked on the quasars;
the uncertainty in the amplitude calibration is 20\%.
Absolute positions in our map have uncertainty due to the
uncertainty in the antenna baselines and the statistical uncertainty 
from the signal-to-noise of the observation. These two 
factors add in quadrature to give an absolute
positional uncertainty of $0\farcs$14.
The lower resolution data (acquired on October 3, 1996, February 2, 1997, and
March 8, 1997) used 0530+135 to track phase variations
and Mars for amplitude calibration.

The L1551~IRS5 data were imaged in four ways which 
stress structures present on different spatial scales.
Figure 1 shows four maps: two with 
robust weightings of the visibilities (robust~=~0.5 yielding
a 3$\farcs 25 \times 3\farcs 04$ beam and
robust~=~-0.25 yielding a 1$\farcs 11 \times 0\farcs 84$ beam),
one with natural weighting of only the high resolution A array 
data restored with the fitted
``clean'' beam (0$\farcs 73 \times 0\farcs 31$ beam),
and one with the A array data restored with a 
circular 0$\farcs$31 ``clean'' beam. 
The latter technique
strongly emphasizes the high resolution information present in the 
A array {\it u,v} data.
With maximum projected baselines ranging from 320~k$\lambda$
to 480~k$\lambda$, the smallest fringe spacings in our dataset
ranges from 0$\farcs$64 to
0$\farcs$44; hence information down to size scales of
0$\farcs$2 to 0$\farcs$3 is present in the {\it u,v} data.
High resolution maps of the secondary quasar 0449+113, 
using the standard technique and using the 0$\farcs$31 ``clean'' beam,
were consistent with a point source.

\section{Results}

Figure 1a (3$\arcsec$ resolution) has a peak flux of 
122$\pm$3~mJy~beam$^{-1}$, and the integrated flux in a 
8$\arcsec$ box centered on the source is 162$\pm$6~mJy.
A Gaussian fit to the image
gives a deconvolved Gaussian source size of 
1$\farcs78 \times 1\farcs75$ and PA=68$^{\arcdeg}$.
Figure 1b (1$\arcsec$ resolution) has a peak flux of
78$\pm$3~mJy~beam$^{-1}$, and the integrated flux in a
3$\arcsec$ box centered on the source is 143$\pm$10~mJy.
A Gaussian fit to the image gives a deconvolved Gaussian source size of
0$\farcs92 \times 0\farcs61$ and PA=157$^{\arcdeg}$.
Figure 1c shows the map of the A array data alone restored with the 
Gaussian fitted clean beam.
The peak flux in the map is 45$\pm$5 mJy beam$^{-1}$, and the
integrated flux in a 1$\farcs$3 box centered
on the source is 75$\pm$11~mJy.
Although it is not obvious in Figure 1c,
over $\frac{1}{2}$ of the flux present in the lowest resolution map
is now gone and the peak flux is roughly $\frac{1}{3}$ of that 
in Figure 1a.
Despite the elongated ``clean'' beam,
the remaining emission is clearly extended north-south 
in the CLEANed image;
a Gaussian fit to the image gives a deconvolved Gaussian source size
of 0$\farcs53 \times 0\farcs$32 and PA=$7.2^{\arcdeg}$.
Figure 1d shows the A array data restored with
the circular $0\farcs 31$ beam. The north-south extension is obvious
in this map and there is no hint of east-west extension. 
The peak flux is 38 mJy beam$^{-1}$ corresponding to
a brightness temperature of 41~K.

The images in Figure 1a and 1b emphasize the overall emission from
the L1551 system.
The reconstructions in Figure 1c and Figure 1d highlight
the small scale emission which is more compact than 
expected for the disk size estimates of Keene and Masson\markcite{kandm90}
(1990) and \markcite{lay}Lay \etal\ (1994).
The compact emission is consistent with arising from two point 
sources, as seen at $\lambda$~=~2~cm and $\lambda$~=~1.3~cm
\markcite{rod86}\markcite{koerner}
(Rodr\'iguez \etal\ 1986; Koerner \& Sargent 1997).
A two Gaussian fit to the $\lambda$~=~2.7~mm emission
in Figure 1d yields the following positions
hereafter labeled IRS5~A and IRS5~B:
IRS5~A:~$\alpha$(J2000) = $04^{h}31^{m}34^{s}.143$,
$\delta$(J2000) = $18^{\arcdeg}08^{\arcmin}05\farcs 09$ and
IRS5~B:~$\alpha$(J2000) = $04^{h}31^{m}34^{s}.141$,
$\delta$(J2000) = $18^{\arcdeg}08^{\arcmin}04\farcs 74$.
These positions agree to within 0$\farcs$05 with the
$\lambda$~=~1.3~cm
source positions of Koerner and Sargent (1997).
The separation of the two sources is 0$\farcs$35, 
corresponding to 49 AU.
Both sources have deconvolved sizes of $\le 0\farcs3$.  
A two point source fit yields flux densities of
45$\pm$6 mJy for IRS5~A and 23$\pm$6 mJy for IRS5~B.
The total flux density in the compact sources is then 68$\pm$9 mJy.

\section{Comparisons with Centimeter High Resolution Data}

High resolution centimeter wavelength images of L1551~IRS5
show two point-like sources and
an extended jet (Bieging \& Cohen 1985). The jet is detected only at
long centimeter wavelengths; the two point sources dominate the
flux at shorter wavelengths.  
The $\lambda$~=~2~cm flux densities are 
1.2 mJy for IRS5~A and 0.93 mJy for IRS5~B (Rodr\'iguez \etal\ 1986).
Recent VLA observations also resolved the
two sources at $\lambda$~=~1.3~cm (Koerner \& Sargent 1997) and yielded
flux densities of 2.0$\pm$0.2~mJy and 1.5$\pm$0.2~mJy, respectively.
The spectral indices between $\lambda$~=~2.0 and 1.3~cm are then 
$\alpha_{A} \sim$1.25 and $\alpha_{B} \sim$1.04, consistent
with $\alpha \sim$1 estimated by Bieging and Cohen (1985).
Extrapolating to 109 GHz, this emission could contribute as much as
$\sim$14.4~mJy and $\sim$7.8~mJy, respectively, to the observed
fluxes.  Hence, the $\lambda$~=~2.7~mm flux is dominated by dust emission.

The proposal of Rodr\'iguez \etal\ (1986)
that the $\lambda$~=~2~cm emission traces the ionized
inner edge of a larger dusty torus is not consistent with the observed
compact $\lambda$~=~2.7~mm emission.  
Since the millimeter emission directly probes the dust,
we should easily see the torus in our high resolution maps.
If there were a torus, the $\lambda$~=~2.7~mm emission would
extend beyond the $\lambda$~=~2~cm sources and, in fact, peak outside
of them.
The binary interpretation of Bieging \& Cohen is consistent 
with our image if the $\lambda$~=~2.7~mm emission 
arises from circumstellar disks within the binary system, while
the $\lambda$~=~2~cm emission traces ionized gas associated with
stellar winds or jets.

\section{The Structure of the L1551~IRS5 System}

Combining all observations to date, 
the L1551~IRS5 system consists of three main circumstellar components:
a large-scale envelope (Keene and Masson 1990; Ladd \etal\ 1995),
a disk or extended structure with a size scale of $\sim$1\arcsec~(Lay 
\etal\ 1994; Keene and Masson 1990), 
and an inner binary system as argued in section 4.
How do these components fit together?
To answer this question, 
\newpage
\noindent we compare our {\it u,v} data 
binned in annuli with simulated observations of models for 
the system, 
binned similarly.
In the following subsections we
discuss each component and derive characteristic masses.

\subsection{Binary Circumstellar Disks}

Figure 2a compares our {\it u,v} data with the Gaussian model 
from Lay \etal\ (1994) scaled to
match the $\lambda$~=~2.7~mm flux at 50~k$\lambda$.  Above
100~k$\lambda$, the Gaussian model is resolved out and does not
fit the data; below 20~k$\lambda$ the data diverge from the model 
due to flux from the envelope.
Figure 2b shows a two point source model with the separation
and amplitudes given in section 3.  The two point sources
beat together to cause the variations in flux seen past 100k$\lambda$.
The proposed binary disk system is evident only in our data; its
separation is too small to be resolved in the data of Lay \etal\ (1994) or 
Keene and Masson (1990).
In fact, due to the small angular size and the embedded nature of the
binary system, the properties of the proposed disks are poorly constrained
by observations to date. The projected separation and
extent of the $\lambda$~=~2.7~mm emission
suggests a maximum outer radius of 25 AU for the disks.
To estimate the masses of the disks, we assume
a standard power-law disk with parameters
characteristic of the HL~Tau disk,
T$_{disk}=330 ({1 AU \over r})^{0.5}$ and
$\Sigma_{disk}\propto r^{-1}$ (Mundy \etal\ 1996;
Beckwith \& Sargent 1991). For dust properties,
we adopt $\kappa$=0.1$({\nu \over 1200\ GHz})$ cm$^2$~g$^{-1}$,
which is consistent with other recent works (e.g.
\markcite{osterloh}Osterloh and Beckwith 1995;
\markcite{ohashi}Ohashi \etal\ 1991;
\markcite{beck}Beckwith \& Sargent 1991).
With these assumptions, the disk masses are
M$_{A} \sim$0.024~M$_{\sun}$ and M$_{B} \sim$0.009~M$_{\sun}$.

\subsection{The Envelope}

The excess emission in our robust weighted maps (Figures 1a and 1b)
compared to our highest resolution map (Figure 1d)
and the rise in flux on baselines shorter than
15 k$\lambda$ (Figure 2), are due primarily to the 
extended envelope. 
Our flux densities in the larger beams are consistent
with previous measurements at similar resolutions:
Keene and Masson (1990) find a peak flux of 130 mJy
beam$^{-1}$ at $\lambda$~=~2.73~mm in a 2$\farcs$6 beam and a total
flux of 150 mJy; \markcite{ohashi}Ohashi \etal\  (1996) measure a total flux
of 160 mJy at $\lambda$~=~2.73~mm using a 4$\farcs$5 beam. 
Our {\it u,v} data in Figure 2 and Figure 2
of Keene and Masson (1990), show similar fluxes around 10~k$\lambda$,
but our data has 20\% to 30\% less flux from 40~k$\lambda$ 
to 70~k$\lambda$.  These differences are within the calibration
uncertainties.

The differences
in flux densities at different resolutions, or equivalently the drop in
flux density with {\it u,v} distance, 
can be used to estimate the properties of the envelope.
Our data are broadly consistent with the envelope
parameters determined by Ladd \etal (1995) and Fuller \etal (1995). Fitting
the drop in flux between 2.6~k$\lambda$ and 15~k$\lambda$ with a
power-law envelope model 
($\rho(r)\propto r^{-1.5}$ and $T(r)\propto r^{-0.5}$)
combined with the two point source model from section 5.1,
reasonable results are obtained for an envelope mass of
$\sim$0.44 M$_{\sun}$, an outer radius of $\sim$1300 AU,
and an inner envelope radius of 30 AU (Figure 2c). 
Steeper envelope density laws ($\rho(r)\propto r^{-2}$) also fit the 
data with a characteristic mass and outer radius of
0.43~M$_{\sun}$ and 1800~AU, respectively.

\subsection{The Circumbinary Structure}

Finally, an intermediate-sized structure, perhaps a circumbinary disk
such as seen around GG~Tau
\markcite{dutrey}(Dutrey, Guilloteau, Simon 1994)
or a ``pseudo-disk''
\markcite{galli}(Galli and Shu 1993),
is needed to
account for the structure resolved by Lay \etal\ (1994) and the
compact structure deduced by Keene and Masson (1990). In our data, this
structure is evident as the excess emission between 30~k$\lambda$ and
90~k$\lambda$ in Figure 2c.  As shown in Figure 2d, this excess can
be fitted with a Gaussian model consistent with that
of Lay \etal\ (1$\farcs$2~$\times$~0$\farcs$7 PA~=~160$^{\arcdeg}$)
 with a flux of 30 mJy plus an envelope model with a mass of 0.28~M$_{\sun}$ 
($\rho(r)\propto r^{-1.5}$) and a radius of 1100~AU.
The parameters of the envelope and the
circumbinary structure are interdependent and hence only crudely
determined.
If the circumbinary structure has dust properties similar to 
the envelope parameters in section 5.2,
the circumbinary structure has a rough mass of 0.04~M$_{\sun}$.

To test the consistency of the above model with the Lay \etal\ (1994)
data, we fit two different source structures to their 
$\lambda$~=~870~$\mu$m data:
a single elliptical Gaussian (a single circumstellar disk) 
and a single elliptical Gaussian with two central point sources
(a circumbinary disk with two small circumstellar disks), 
following the fitting procedure of
\markcite{lay_dis}Lay \etal\ (1994; also see Lay 1994).
The model did not include envelope emission
since the JCMT-CSO baselines ranged from 50~k$\lambda$ to 200~k$\lambda$,
where the envelope emission is completely resolved out.
The single elliptical Gaussian model fits the $\lambda$~=~870~$\mu$m
data very well, with parameters comparable to those found 
by Lay \etal\ (1994). 
The addition of two point sources to the single Gaussian model
produces as good a fit as the single Gaussian model, but the FWHM
of the Gaussian increases slightly.  Hence, the data cannot
distinguish between the single Gaussian and single Gaussian
with point source models.
If the circumbinary material is optically thick
at $\lambda$~=~870~$\mu$m, the 
Lay \etal\ data would not even see the embedded 
circumstellar disks.
If the circumbinary material is not optically thick, the Lay \etal\
data place a limit on the flux from the circumstellar disks:
at a 95\% confidence level the circumstellar disks emit $\leq$~1.3~Jy
at $\lambda$~=~870~$\mu$m. 

\section{Young Binary systems}

Our data present the first direct detection of a close, embedded
binary system.
Proposed wider binary systems have been
identified among embedded sources, e.g. 
IRAS 16293-2422 (\markcite{wootten}Wootten 1989), 
NGC~1333~IRAS4 (\markcite{sandell}Sandell \etal\ 1991;
\markcite{lay2}Lay \etal\ 1995),
and L1527 (\markcite{fuller96}Fuller, Ladd, \& Hodapp 1996),
but the number of such systems is actually 
quite small compared to the number of known embedded sources.
Surveys of pre-main sequence (PMS) stars
find that binary systems are at least as common among young visible stars
as among main-sequence stars 
(\markcite{simon}Simon \etal\ 1992; 
\markcite{ghez}Ghez, Neugebauer \& Matthews 1993; 
\markcite{leinert}Leinert \etal\ 1993; 
\markcite{reipurth}Reipurth \& Zinnecker 1993); so
binaries should be common among young, deeply embedded systems.
That they have not often been seen is probably due to the lack
of sub-arcsecond resolution observations which are necessary to resolve
close binaries.
The separation of the L1551~IRS5 system is near the median
separation for main sequence binaries ($\sim$30~AU,
\markcite{duq}Duquennoy \& Mayor 1991).
The low detection rate of wide embedded binaries is in rough
agreement with the fraction of main sequence binaries with 
separations between 300 and several 1000 AU.

L1551~IRS5 also ranks as one of the few close binary systems
with significant dust emission associated with both components.
Submillimeter wavelength
surveys have generally found lesser amounts of
dust emission associated with PMS binary systems
than with young single stars systems (\markcite{simon}\markcite{simon95}Simon 
\etal\ 1992, 1995). 
In a statistical comparison of binaries and single stars, 
\markcite{jensen}Jensen, Mathieu \& Fuller (1996) found that
binaries with separations $\le$ 50-100 AU statistically have
lower submillimeter fluxes than wider binaries, but 
wide binaries are indistinguishable from single stars;
hence, the L1551~IRS5 system may be unusual.  However,
these studies concentrate on T Tauri stars and 
exclude the youngest sources, Class I or younger.
It is possible that embedded close binaries, which are still accreting
mass, have substantial circumstellar or circumbinary disks which
disappear later when the envelope is no longer
feeding-in material. 

\section{Conclusions}

Sub-arcsecond $\lambda$~=~2.7~mm observations of L1551~IRS5 have
resolved a compact central structure, which is most plausibly interpreted as
a young binary
system. The $\lambda$~=~2.7~mm continuum emission shows two peaks
which are similar, in absolute position and separation, to 
the free-free emission observed at centimeter wavelengths.
Our interpretation is that we are
detecting thermal dust emission from small disks around the individual stars
in a binary system and that the centimeter emission arises in the 
associated stellar winds.
We propose that the L1551~IRS5 system is composed
of two circumstellar disks, located inside
a circumbinary structure, embedded in a large-scale envelope.
Simple modeling yields masses for these components:
circumstellar disk masses of 0.024~M$_{\sun}$
and 0.009~M$_{\sun}$ for the northern and southern sources respectively,
a circumbinary structure mass of 0.04~M$_{\sun}$, and an envelope mass
of 0.28~M$_{\sun}$.
The binary separation for L1551~IRS5 is about 50 AU, close to the median
separation for main sequence binaries.  
The small number of young embedded binaries detected to date,
probably reflects the inadequate 
angular resolution available in the earlier studies, rather
than an intrinsic sparsity of binaries.

\acknowledgments
We thank the Hat Creek staff for their efforts in the construction
and operation of the long baselines array.
We also thank the referee, Jocelyn Keene, and Oliver Lay for their comments.
This work was supported by NSF Grants NSF-FD93-20238 and AST-9314847. LGM
acknowledges support from NASA grant NAGW-3066.

\newpage

\begin{figure}
\includegraphics{looney.et.al.figures/figure1.ps}
\vspace{4.5in}
\caption{
$\lambda$~=~2.7~mm maps of the continuum emission from
L1551~IRS5. 
a) Robust weighting of 0.5 map made with 
data from three arrays.
The beam is 3$\farcs25 \times 3\farcs04$ PA~=~29$^{\arcdeg}$,
and the RMS noise is 2.5 mJy beam$^{-1}$.  The contours are 
-3,-2,2,3,4,5,6,7,8,9,10,15,20,25,30,35 times 3.3~mJy beam$^{-1}$ 
(the RMS from Panel b).
b) Robust weighting of -0.25 map made with
data from three arrays.
The beam is 1$\farcs11 \times 0\farcs84$ PA~=~60$^{\arcdeg}$,
and the RMS noise is 3.3~mJy beam$^{-1}$.  The contours are
the same as in Panel a.
c) Naturally weighted map made from only the A array data.
The beam is 0$\farcs 73 \times 0\farcs 31$ PA=47$^{\arcdeg}$
and the RMS is 4.5~mJy beam$^{-1}$.
The contours are in steps of 1 $\sigma$
starting at $\pm$2~$\sigma$.
d) A array naturally weighted data,
restored with a circular 0$\farcs$31 beam.
The contours and RMS are the same as in Panel c.
The two crosses in Panels c and d mark the 
$\lambda$~=~1.3~cm source positions from Koerner \&
Sargent 1997.
The restoring beam in each panel is shown in the lower
left-hand corner.}
\end{figure}

\newpage

\begin{figure}
\includegraphics{looney.et.al.figures/figure2.ps}
\vspace{4.5in}
\caption{
The measured $\lambda$~=~2.7~mm visibilities binned in annuli 
(open squares) compared with different model visibilities
(gray, closed squares).
a) The Lay \etal\ (1994) model Gaussian scaled
to match the $\lambda$~=~2.7~mm fluxes around 50~k$\lambda$.
b) Two point source model constrained by
fitting Figure 1d.  
c) Characteristic model fit with an envelope 
(0.44~M$_{\sun}$, $\rho(r)\propto r^{-1.5}$, 
$T(r)\propto r^{-0.5}$, and 1300~AU radius)
and the two point sources from Panel b.
d) Characteristic fit for a model with 
an envelope (0.28~M$_{\sun}$, $\rho(r)\propto r^{-1.5}$,
$T(r)\propto r^{-0.5}$, and 1100~AU radius),
a Gaussian (30 mJy, 1$\farcs2 \times 0\farcs7$ PA~=~160$^{\arcdeg}$),
and the two point sources from Panel b.}
\end{figure}

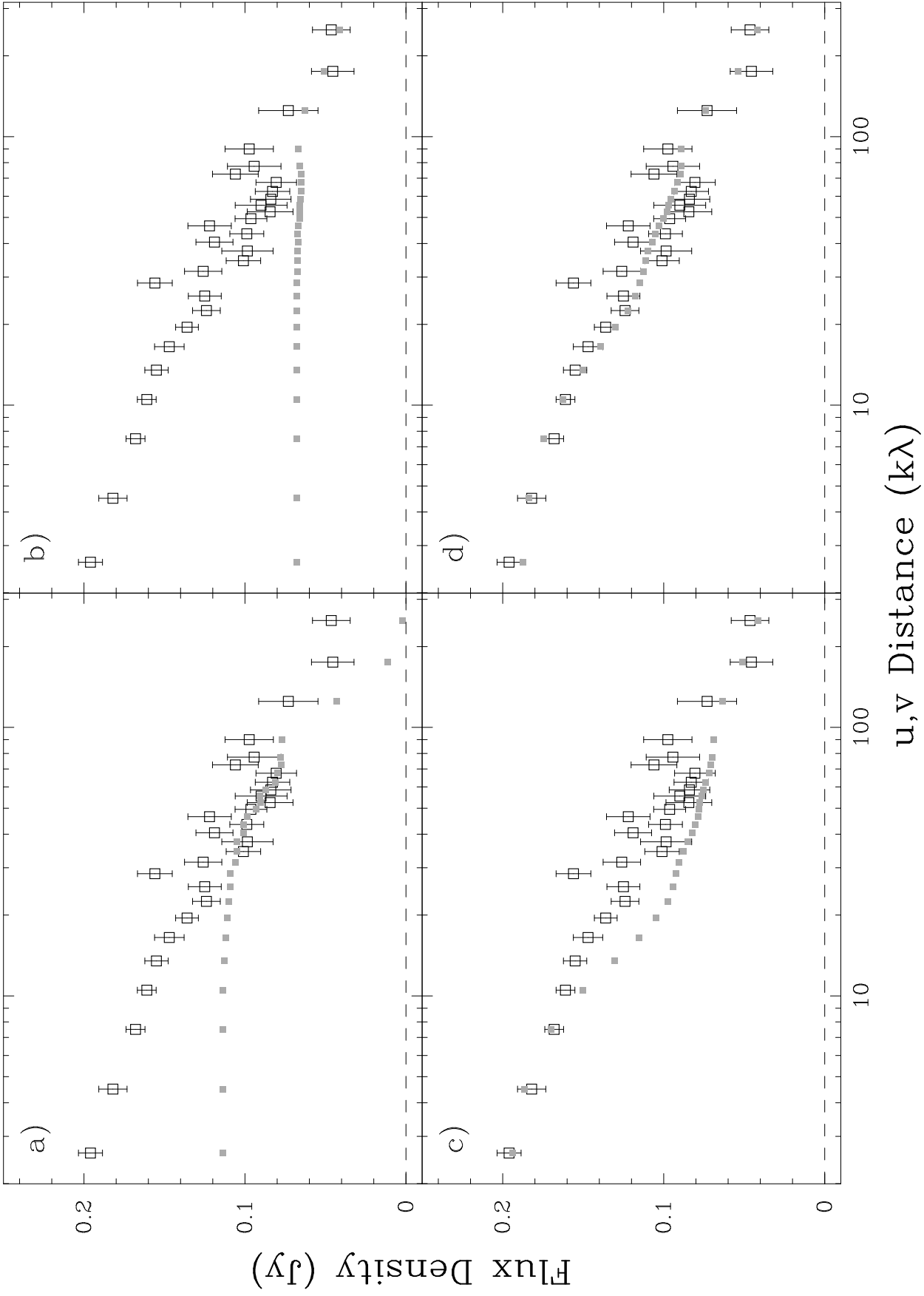
\end{document}